# Enhancing the Open-Circuit Voltage of Perovskite Solar Cells by up to 120 mV using π-Extended Phosphoniumfluorene Electrolytes as Hole Blocking Layers

*Qingzhi An, Qing Sun, Andreas Weu, David Becker-Koch, Fabian Paulus, Sebastian Arndt, Fabian Stuck, A. Stephen K. Hashmi, Nir Tessler and Yana Vaynzof**

Q. An, Q. Sun, A. Weu, D. Becker-Koch, Dr. F. Paulus, Prof. Dr. Y. Vaynzof
Kirchhoff Institute for Physics, Im Neuenheimer Feld 227, Heidelberg University, Germany

Q. An, Q. Sun, A. Weu, D. Becker-Koch, Dr. F. Paulus, Prof. Dr. Y. Vaynzof
Centre for Advanced Materials, Im Neuenheimer Feld 225, Heidelberg University, Germany

Dr. S. Arndt, F. Stuck, Prof. Dr. A. S. K. Hashmi
Institute for Organic Chemsitry, Im Neuenheimer Feld 270, Heidelberg University, Germany

Prof. Dr. A. S. K. Hashmi
Chemistry Department, Faculty of Science, King Abdulaziz University (KAU), 21589 Jeddah, Saudi Arabia

Prof. Nir Tessler
Sara and Moshe Zisapel Nano-Electronic Center, Department of Electrical Engineering, Technion-Israel Institute of Technology, Haifa 32000, Israel

E-mail: vaynzof@uni-heidelberg.de

Four π-extended phosphoniumfluorene electrolytes (π-PFEs) are introduced as hole-blocking layers (HBL) in inverted architecture planar perovskite solar cells (PVSCs) with the structure of ITO/PEDOT:PSS/$MAPbI_3$/PCBM/HBL/Ag. The deep-lying highest occupied molecular orbital (HOMO) energy level of the π-PFEs effectively blocks holes, decreasing contact recombination. We demonstrate that the incorporation of π-PFEs introduces a dipole moment at the PCBM/Ag interface, resulting in a significant enhancement of the built-in potential of the device. This enhancement results in an increase in the open-circuit voltage of the device by up to 120 mV, when compared to the commonly used bathocuproine HBL. The results are confirmed both experimentally and by numerical simulation. Our work demonstrates that interfacial engineering of the transport layer/contact interface by small molecule electrolytes is a promising route to suppress non-radiative recombination in perovskite devices and compensate for a non-ideal energetic alignment at the hole-transport layer/perovskite interface.

## 1. Introduction

Lead halide perovskites are an exciting class of materials that when applied in photovoltaic devices have reached a certified power conversion efficiency (PCE) of 23.7%, demonstrating a great potential to compete with the commercially available crystalline silicon solar cells.[1] While early research efforts focused on the mesoporous structure, in which the perovskite layer is deposited onto a mesoporous inorganic scaffold,[2-3] recently planar perovskite solar cells attracted more research attention due to a simpler manufacturing method and low-processing temperature.[4] Unlike mesoporous devices that exist only in the n-i-p structure, planar perovskite photovoltaic devices can be made in two architectures: the n-i-p standard structure or the inverted p-i-n structure. In the latter, the most common configuration consists of a perovskite active layer that is sandwiched between a hole transport layer (HTL) poly(3,4-ethylenedioxythiophene):poly (styrenesulfonate) (PEDOT:PSS) and electron transport layer (ETL) phenyl-$C_{61}$-butyric acid methyl ester (PCBM). Both transport layers can be easily processed from solution at low temperatures. PEDOT:PSS is a well-established HTL commonly used also in organic photovoltaic devices due to its high conductivity and low absorption in the visible range.[5] However, due to the mismatch in energy levels between the PEDOT:PSS and the perovskite layers, the PEDOT:PSS/perovskite interface introduces significant energy losses, resulting in a lowering of the open-circuit voltage ($V_{OC}$) of the device.[6] To resolve this issue, various strategies have been suggested in order to enhance the open-circuit voltage in inverted architecture perovskite solar cells, including doping the PEDOT:PSS layer in order to deepen its work function,[7-9] replacing it with other high work function HTLs[10-11] and changing the fabrication method of the perovskite layer in order to enhance its quality.[12-13]

The use of PCBM as ETL is motivated by the negligible J-V hysteresis associated with trap states passivation by PCBM at the perovskite/PCBM interface.[14] While efficient electron extraction from the perovskite layer has been demonstrated for PCBM and other fullerenes,[15-16] the relatively low ionization potential of PCBM results in poor hole-blocking and a significant charge recombination, resulting in a poor device performance with a particularly low fill factor.[17] Additionally, the mismatch in the energetic alignment between the lowest unoccupied molecular orbital (LUMO) of PCBM and the cathode work function results in a further reduction of photovoltaic performance. To address these two issues, a range of materials have been investigated as modifiers at the PCBM/cathode interface. For examples, interlayers of N,N'-bis(1-n-hexylpyridin-1-ium-4-ylmethyl)-1,4,5,8-naphthalenetetracarboxydiimide (PN6),[18] ethoxylated polyethylenimine (PEIE),[19] (11-mecaptoundecyl)-trimethylammonium bromide (MUTAB),[20] amino-functionalized polymer $PN_4N$,[21] metal acetylacetonate[22] and Rhodamine[23] were reported to modify the cathode work function, greatly improving the device efficiency. Materials such as ZnO nanoparticles,[24] $TiO_x$,[25] bathocuproine (BCP)[26] and bispyridinium salt FPyBr[27] were also utilized aiming to block holes and suppress charge loss. With many of these materials, the improvements in device performance originate mostly from enhancement in the fill factor, especially for devices that utilize PEDOT:PSS as a HTL. In these devices the open-circuit voltages, while improved when compared to reference device without a HBL, remain in the range of 0.9V to 1V,[13,19-21,24] lagging behind other HTL such as NiOx, poly[bis(4-phenyl)(2,4,6-trimethylphenyl)amine] (PTAA) or poly[N,N'-bis(4-butylphenyl)-N,N′ -bis(phenyl)benzidine] (poly-TPD).[28-30]

Herein, we report on the application of a series of π-extended phosphoniumfluorenes electrolytes (π-PFE) as hole-blocking layers photovoltaic devices with the structure: ITO/PEDOT:PSS/$MAPbI_3$/PCBM/HBL/Ag. The high ionization potential of π-PFEs can efficiently block holes from the cathode, greatly reducing charge recombination losses. Most importantly, π-PFEs result in the formation of a dipole at the interface with the cathode Ag, significantly increasing the built-in field of the devices. As a consequence, devices with π-PFEs exhibit an enhanced $V_{OC}$ of up to 1.07 V, expressively surpassing those with the commonly used BCP (~0.95 V) resulting in a maximum PCE of 18.46%.

## 2. Results and Discussion

### 2.1. Photovoltaic performance of PVSCs with π-PFE HBLs

The photovoltaic device structure employed in this work is schematically illustrated in **Figure 1a**. Devices with no HBL, a commonly used BCP or the π-PFE electrolytes HBL were fabricated with all remaining layers kept unchanged. The chemical structures of the π-extended phosphoniumfluorenes (π-PFE1-4) are shown in **Figure 1b**. To identify whether π-PFE1-4 can serve as efficient HBLs, their ionisation potential (IP) was measured using ultra-violet photoemission spectroscopy (**Figure 1c**). As reference, the IP of PCBM and BCP/PCBM was also measured. The measurements show that both the BCP and the π-PFEs layers exhibit a deep HOMO (IP = 6.8-7 eV) resulting in the formation of a large hole injection barrier from the underlying PCBM. Based on these results, both BCP and π-PFEs should efficiently block holes from reaching the Ag electrode, which is not expected in the case of a device with no HBL due to the relatively low IP of PCBM (IP = 5.85 eV).

While the IP measurements of the π-PFEs suggest that incorporating π-PFEs into the device structure should have a very similar effect on the device performance to that of BCP, electrical characterization reveals that this is not the case. **Figure 2a** shows the current density-voltage (J-V) curves of photovoltaic devices measured in the dark, demonstrating that the 'knee' position of the diode characteristics is significantly shifted to higher voltages in devices with π-PFEs HBLs. Such a shift corresponds to an enhanced built-in potential and could lead to an increased $V_{OC}$ of the devices upon illumination. J-V measurements acquired under AM 1.5 G simulated 100 mW/cm$^2$ sunlight (**Figure 2b**) show the significant changes in the photovoltaic performance of the devices. As expected, the bare PCBM control device exhibits a low fill factor (FF) of 60.1 % and a moderate open circuit voltage ($V_{OC}$) of 0.96 V, leading to a power conversion efficiency (PCE) of only 12.85 %. The J-V characteristics of this device shows an S-shape originating from a high series resistance and poor hole-blocking, in agreement with previous reports.[19,31] The introduction of any HBL (BCP/π-PFEs) eliminates the S-shape, significantly increasing the FF of the devices. This increase in FF is consistent with the IP measurements above, as the deep lying HOMO level of π-PFEs and BCP can efficiently prevent the electron-hole recombination between PCBM and silver contact, resulting in the improvement of FF. However, while the commonly used BCP has little to no effect on the device $V_{OC}$, the use of π-PFEs results in a significant increase by as much as 120 mV. The short-circuit current of the devices remains very similar and is not affected by the introduction of a HBL. Overall, the improvements in FF and $V_{OC}$ result in a maximum PCE of 18.46% for a device with a π-PFE HBL, surpassing that of the BCP reference (17.38%). We note that in order to ensure that the comparison is performed to the best possible devices with a BCP hole-blocking layer, we optimized the thickness of the BCP layer in our solar cells (**Figure S1**). We find that the optimal thickness of the BCP layer is 5 nm, in agreement with previous work by Chen *et al.*[32] Consequently, all BCP devices discussed in this work employ this optimal layer thickness. The photovoltaic parameters for both reverse and forward scans are summarized in **Table 1** and a complete statistic over ~260 devices is shown in Supplementary Information, **Figure S2**. The results show that the choice of side chain (1-4) has no significant effect on the photovoltaic performance.

The external quantum efficiency (EQE) spectra of the devices are shown in **Figure 2c**. The EQE spectra are very similar, in agreement with the similarly measured $J_{SC}$ of devices with and without the HBLs. This similarity indicates that upon integrating π-PFE1-4 (or BCP) into the device structure, the charge carrier extraction efficiency at the PCBM/Ag interface remains largely unaffected. **Figure 2d** shows the time dependence of the maximum power output efficiency of the devices, with all devices exhibiting a stable output efficiency with a low hysteresis, as is common for devices with a PCBM electron extraction layer.[14]

## 2.2 Non-radiative recombination processes in PVSCs with π-PFE HBLs

To investigate in depth the origin of the increase in $V_{OC}$ for devices with π-PFE HBLs, we characterized their effect on the charge-carrier radiative recombination processes in the device. Time correlated single photon counting (TCSPC) measurements (**Figure 3a**) were collected on a $MAPbI_3$ perovskite only film, as well as on PCBM/$MAPbI_3$ and HBL/PCBM/$MAPbI_3$. The bare reference $MAPbI_3$ film shows an almost mono-exponential decay with a long lifetime of 552 ns. Upon coating the perovskite film with PCBM, an additional, faster decay appears with a lifetime of only 1.23 ns due to efficient electron transfer from $MAPbI_3$ to PCBM. Further quenching is observed in the case of the BCP/PCBM/$MAPbI_3$ sample with an even shorter lifetime of 0.8 ns. In contrast, depositing π-PFE1-4 on PCBM/$MAPbI_3$ slows down the photon decay slightly increasing fluorescence lifetimes to approximately 3 ns. This increase suggests that the use of π-PFE HBLs suppresses non-radiative recombination processes, which would correspond to an increase in the device $V_{OC}$. A decrease in the non-radiative recombination should also manifest itself as an enhancement of the electroluminescence of the devices. The electroluminescence quantum efficiencies (ELQE) as a function of current density of perovskite solar cells with different HBLs are shown in **Figure 3b**. The ELQE of devices with a BCP HBL is very similar to that of the reference PCBM-only device. However, the incorporation of π-PFE HBLs results in an increase of approximately one order of magnitude in the device ELQE. This increase in ELQE is in agreement with the previously observed increase in the $V_{OC}$ of the devices, as expected from Rau's reciprocity relation.[33]

To further investigate the charge recombination processes in the PVSC devices, we performed transient photo-voltage (TPV) and light intensity-dependent open-circuit voltage measurements.

In the TPV studies, charges were generated by exposing the devices (held under open circuit conditions with a white light bias) to a weak laser pulse and the transient photo-voltage associated with the charge population perturbation was tracked to investigate the charge recombination processes. As shown in **Figure 4a**, solar cells with π-PFE HBLs exhibited a longer charge recombination lifetime than those with either bare PCBM or BCP-based devices, indicating that charge recombination is suppressed by incorporation of π-PFE HBLs. Light intensity dependent $V_{OC}$ measurements (**Figure 4b**) show that devices with π-PFE HBLs exhibit an overall lower trap-assisted recombination, as the slopes measured for bare PCBM, PCBM/BCP and PCBM/π-PFE based devices were calculated to be 1.35 $K_BT/q$, 1.34 $K_BT/q$ and 1.23±0.04 $K_BT/q$, respectively. This suggests that while trap-assisted charge recombination is present in all the devices, it is the lowest in devices that incorporate π-PFE HBLs.[34]

Since recombination pathways at the perovskite/PCBM interface remain unchanged with the different HBLs, the reduction in trap-assisted recombination must be associated with the PCBM layer or the interface PCBM/HBL/Ag. Since it was previously reported that enhancing the order of the PCBM electron transport layer leads to a higher $V_{OC}$ in perovskite solar cells,[35] we investigated the PCBM layers with and without the HBL using X-ray Diffraction (Supplementary Information, **Figure S3**) and photothermal deflection spectroscopy (Supplementary Information, **Figure S4**), which can probe the crystalline and energetic order in the film, respectively. The measurements showed no significant difference between the samples, suggesting that within the experimental resolution, the PCBM film remained unaffected by the deposition of the π-PFE HBLs.

To further understand the changes induced by the incorporation of π-PFE HBLs, we performed charge extraction measurements as previously applied to organic devices.[36] In short, these measurements allow quantifying the amount of extractable charge in the device at a given voltage, thus providing information about the available density of states and built-in field. Comparing the measurements performed on PCBM and PCBM/BCP devices with those that incorporate π-PFE HBLs (**Figure 4c**) shows at the same charge carrier concentration (so same quasi Fermi levels separation), the voltage obtained for the π-PFE HBLs devices is significantly higher. This observation may arise from either a significant change in the density of states (reduced energetic disorder) or an increase in the built-in potential of the device. As significant changes to the density of states have already been ruled out, the increased voltage originates predominately from the increase in the built-in potential across the device. Conductivity measurements performed on unipolar diodes with the structure ITO/ZnO/PCBM/HBL/Ag (**Figure 4d**), which show a minor increase (on average fivefold) in charge carrier mobility and overall current for devices with π-PFE HBLs (Supplementary Information, **Table S1**) as compared to those with BCP. We hypothesise that this increase is associated with the effect of Ag electrode evaporation on the trap density in PCBM. It has been demonstrated that metal evaporation onto PCBM results in extensive penetration of the metal into the bulk of the organic layer.[37] The presence of a HBL layer may have an effect on the penetration of Ag atoms into PCBM. In the case of BCP, it is known that BCP-metal complexes are formed upon the evaporation of the cathode, suggesting that Ag is able to penetrate thin BCP layers as those are likely to exhibit similar mechanical properties as PCBM[38] and must be kept very thin (~5 nm) to allow for optimum device operation (**Figure S1**).[32] Our measurements suggest that the average thickness of the π-PFE HBLs is ~10 nm, which is likely to lead to a reduced penetration of the metal atoms into the PCBM without compromising the photovoltaic performance of the device. Additionally, the ionic character of the electrolyte HBL is likely to further suppress the penetration of Ag into the PCBM layer. An alternative explanation could be related to an n-doping of PCBM by the anions of the π-extended phosphoniumfluorenes,[27] which would also lead to an enhanced mobility. We note that such a minor increase in mobility further confirms that no significant changes to the energetic disorder occur, as an average fivefold increase would correspond to a reduction in the energetic disorder on the other of only 0.5kT,[39] insignificant when compared to the enhancement in $V_{OC}$ we observe.

## 2.3 Energetic alignment at the PCBM/π-PFE/Ag interface

To elucidate the origin of the increase built-in potential, we focused on the investigation of the interface with the Ag cathode. Based on extensive research of both small molecule electrolytes and polyelectrolytes as modifiers in organic photovoltaic devices,[40-41] it is reasonable to expect that their application as a HBL may induce a dipole at the PCBM/Ag interface. To probe the energy level alignment of at the PCBM/π-PFE1-4/Ag interface, we carried out UPS measurement on reference Ag, BCP/Ag and π-PFE1-4/Ag layers that were detached from ITO/PCBM/HBL/Ag devices, similar to the approach used by Lee *et al.*[42] As shown in **Figure 5a**, the photoemission onset of π-PFE1-4/Ag samples shifts by approximately 0.4 eV, corresponding to a shift in the vacuum level upwards at the π-PFE1-4/Ag interface (**Figure 5b**). No such shift is observed in the case of BCP/Ag. These measurements suggest that a dipole is formed at the interface of π-PFE1-4/Ag which is associated with the N- anions being located near the Ag contact, with the cations (P+) remaining in the proximity of the PCBM surface (**Figure 5c**). The direction of the dipole formed at this interface will lead to an enhancement in the built-in potential of the device and subsequently, its open-circuit voltage. We note that in the case of organic photovoltaics, it has also been shown that a post processing treatment with a methanol (which has been used as the solvent for the deposition of π-PFE1-4 layers) may lead to an increase in the PEDOT:PSS work function and enhancement in the built-in potential and $V_{OC}$,[43] but our reference measurements show that this is not the case for perovskite solar cells (Supplementary Information, **Figure S5**)

To explore whether the positive enhancement may be achieved also for other, higher work function hole transporting layers than PEDOT:PSS, we fabricated and characterised devices with the structure ITO/PTAA/$MAPbI_3$/PCBM/π-PFE1-4/Ag (Supplementary Information

**Figure S6** and **Table S2**). Reference devices with BCP HBL reach impressive $V_{OC}$ of ~1.1 V, in agreement with previous reports.[29] The incorporation π-PFE1-4 HBL still results in an increase in $V_{OC}$ by up to 30-40 mV - a less pronounced increase than in the case of PEDOT:PSS. This result is in agreement with the expected higher built-in potential of the PTAA based devices even when including BCP as a HBL due to the high work function of PTAA. The difference in the magnitude of the effect of π-PFEs when applied to devices with PEDOT:PSS and PTAA hole transport layers also suggests that electrolyte HBLs can be strategically used to compensate for the unfavorable energetic alignment at the HTL/perovskite interface, such as in the case of PEDOT:PSS/$MAPbI_3$.

### 2.4 Numerical simulations of PVSCs with different HBLs

To confirm that the enhancements in the photovoltaic performance originate from the introduction of the dipole at the interface with Ag, we performed numerical simulations based on the same approach described in 44. In short, we numerically solve the drift diffusion and Poisson equation self-consistently for electrons, holes and ions. The ions exist and are mobile only within the perovskite active layer, while the electrons and holes are free to move also within the transporting/blocking layers. To produce the J-V curves under one sun illumination, we solve at each bias point for the steady state of both ions and electronic charges. This would mimic a very slow scan that cancels the small hysteresis and allows to focus on the shape of the J-V curves.

The material parameters used in the simulation are listed in the Supplementary Information, **Table S3**. To incorporate all the experimental observations into the model, we increased the electron mobility of the PCBM layer by a factor of five in the case of π-PFE, as the average enhancement observed for the various electrolyte HBLs (Supplementary Information, **Table S1**). We have also included a 0.4 eV dipole across the 10 nm thick π-PFE HBL. Further details concerning the simulation can be found in the Supplementary Information, **Supp. Note 1**.

**Figure 6** shows the simulated J-V curves for devices with either no HBL or BCP and π-PFE HBL. Similar to the experimental results, strong S-shape is observed for devices that do not include a HBL. The incorporation of BCP eliminates the S-shape, but does not result in an increase in the $V_{OC}$, in agreement with the experimental results. Replacing the BCP with a π-PFE HBL results in a significant enhancement in the $V_{OC}$ on the order of 100 mV – in excellent agreement with the average enhancement observed for π-PFE1-4. We note that including only one of the positive effects of π-PFE (i.e. the enhanced mobility or the interfacial dipole) did not result in a comparable increase in $V_{OC}$, further confirming that both effects contribute to the observed improvement.

## 3. Conclusion

In summary, the application of π-extended phosphoniumfluorene electrolytes as hole blocking layers in perovskite solar cells have been demonstrated to enhance device $V_{OC}$ due to their ability to efficiently suppress non-radiative recombination processes and form a dipole moment between electron extraction layer and cathode contact. Consequently, the $V_{OC}$ can be improved by up to 120 mV resulting in a maximum power conversion efficiency of 18.5 %, surpassing the performance of commonly used bathocuproine hole blocking layer (17.4 %). Our work highlights the tremendous potential of electrolyte materials as hole blocking layers in perovskite photovoltaics and shows that interfacial engineering efforts should not be limited to the interfaces with the perovskite layer, but extended also to the optimization of the energetic alignment at the contacts.

## 4. Experimental Section

*Materials:* Methylammonium iodide ($CH_3NH_3I$) was purchased from GreatCell Solar. PEDOT: PSS was purchased from Heraeus Deutschland GmbH&Co. PCBM (99.5%) were purchased from Solenne BV. π-extended phosphoniumfluorenes were synthesized based on the previous work.[45] All other materials were purchased from Sigma-Aldrich and used as received.

*Photovoltaic Device Fabrication:* Pre-patterned indium tin oxide (ITO) coated glass substrates (PsiOTech Ltd., 15 Ohm/sqr) were ultrasonically cleaned with 2 % hellmanex detergent, deionized water, acetone, and isopropanol, followed by 8 min oxygen plasma treatment. PEDOT:PSS prepared based on the previous report[8] was spun coat on the clean substrates with 4000 rpm 30 s and annealed at 150 °C 15 min. A lead acetate trihydrate $MAPbI_3$ recipe[16] was used for forming the $MAPbI_3$ perovskite layer, in detail the perovskite solution was spin coated at 2000 rpm for 60 s in a dry air filled glovebox (RH < 0.5 %). After blowing 25 s and drying 5 min, the as-spun films were annealed at 100 °C for 5 min forming a black and uniform perovskite layer. The prepared samples were transferred to a nitrogen filled glove box, PCBM (20 mg/ml dissolved in chlorobenzene) were dynamically spun coat at 2000 rpm 30s on the perovskite layer followed by a 10 min 100 °C annealing. Sequentially BCP (0.5mg/ml dissolved in isopropanol) or the π-extended phosphoniumfluorenes (0.5mg/ml dissolved in methanol) was spun coat on the PCBM, forming a thin layer around 5 nm. To complete the device, 80 nm silver was deposited via thermal evaporation under high vacuum.

*Photovoltaic Device Characterization:* The current density-voltage (J-V) was measured by a computer controlled Keithley 2450 Source Measure Unit under simulated AM 1.5 sunlight with 100 mW $cm^{-2}$ irradiation (Abet Sun 3000 Class AAA solar simulator). The light intensity was calibrated with a Si reference cell (NIST traceable, VLSI) and corrected by measuring the spectral mismatch between the solar spectrum, the spectral response of the perovskite solar cell and the reference cell. The mismatch factor was calculated to be 10 %. The cells were scanned from forward bias (1.2 V) to short circuit and reverse at a rate of 0.25 V $s^{-1}$ by employing a mask to eliminate the overestimate of the photocurrent. No preconditioning was applied prior to measurements. Light intensity dependence measurements were carried out using neutral density filters.

*Ultra-violet Photoemission Spectroscopy (UPS):* Ultra-violet photoemission spectroscopy measurements were performed on π-PFE/PCBM to characterize their ionization potential and on detached Ag electrodes to probe the effect of π-PFE on the Ag work function. The samples were transferred to an ultrahigh vacuum (UHV) chamber of the PES system (Thermo Scientific ESCALAB 250Xi) for measurements. UPS measurements were carried out using a double-differentially pumped He discharge lamp (hν = 21.22 eV) with a pass energy of 2 eV and a bias of -10 V.

*Photothermal Deflection Spectroscopy (PDS):* PCBM, BCP/PCBM and π-PFE/PCBM layers for PDS characterization were prepared on spectrosils in an identical method to their processing in a device. Under inert conditions the samples were emerged in a signal enhancing liquid (Fluorinert FC-770). The samples were excited using a tunable, chopped, monochromatic light source (150W xenon short arc lamp with a Cornerstone monochromator) and probed using a laser beam (He-Ne laser from REO) propagating parallel to the surface of the sample. The heat generated through the absorption of light changes the reflective index of Fluorinert, resulting in the deflection of the laser beam. This deflection is measured using a position sensitive-detector (Thorlabs, PDP90A) and a lock-in amplifier (Amatec SR7230) and directly correlated to the absorption of the film.

*X-ray Diffraction (XRD):* XRD measurements of the PCBM and PCBM/π-PFE1-4 films on silicon were conducted on a Rigaku SmartLab diffractometer with a 9kW rotating copper anode in Bragg-Brentano geometry. Diffraction patterns (intensity vs. 2θ) were recorded with a HyPix3000 detector operated in 1D-mode equipped with a $k_{\beta}$ filter.

*Time Correlated Single Photon Counting (TCSPC):* $MAPbI_3$, PCBM/$MAPbI_3$, BCP/PCBM/$MAPbI_3$ and π-PFE/PCBM/$MAPbI_3$ layers were measured using a LifeSpec II TCSPC instrument (Edingburgh Instruments). A 475 nm pulsed LED was used as excitation source (pulse width ~750ps) with a repetition rate of 200 kHz and a fluency of 0.38 mJ/cm$^2$. The central wavelength of 775 nm was detected with a bandwidth of 25 nm.

*Transient Photovoltage/Photocurrent:* For transient photovoltage/photocurrent measurements, the light of an inorganic LED (Thorlabs TO-1 ¾, λ = 465 nm) was pulsed by a function generator (Agilent/Keysight 33510B) and focused on the solar cell. An oscilloscope (Picoscope 5443A) with and without a 50 Ω terminator placed across the oscilloscope input was used to measure the transient photocurrent and transient photovoltage, respectively.

*Electroluminescence Measurements:* The devices were characterized in an integrating sphere (Labsphere Inc.). The current−voltage characteristics were measured using a source-measure unit (Keithley 2450). At the same time the emitted light spectra were recorded using a scientific grade spectrometer (Ocean Optics QE65PRO) and converted to luminance. The optical system (integrating sphere, spectrometer, and coupling optical fiber) were calibrated using a calibrated light source (Ocean Optics HL-2000-CAL).

*Unipolar Device Fabrication and Characterization:* Unipolar devices were prepared in the structure: Glass/ITO/ZnO/PCBM/HBL/Ag. The ZnO layer has been deposited following previous work.[46-47] The current−voltage characteristics of the devices were measured using a source-measure unit (Keithley 2450).

*Device Simulation:* A previously developed model that includes the contributions of charges and ions has been used to simulate the influence of incorporating π-PFE HBL into the device structure.[44] Further details are provided in the Supplementary Material.

**Acknowledgements**

This work has received financial support of the Deutsche Forschungsgemeinschaft (DFG) SFB 1249 projects A03 and C04. This project has also received funding from the European Research Council (ERC) under the European Union's Horizon 2020 research and innovation programme (ERC Grant Agreement n° 714067, ENERGYMAPS). We would like to kindly thank Prof. Uwe Bunz for providing access to the device fabrication facilities.

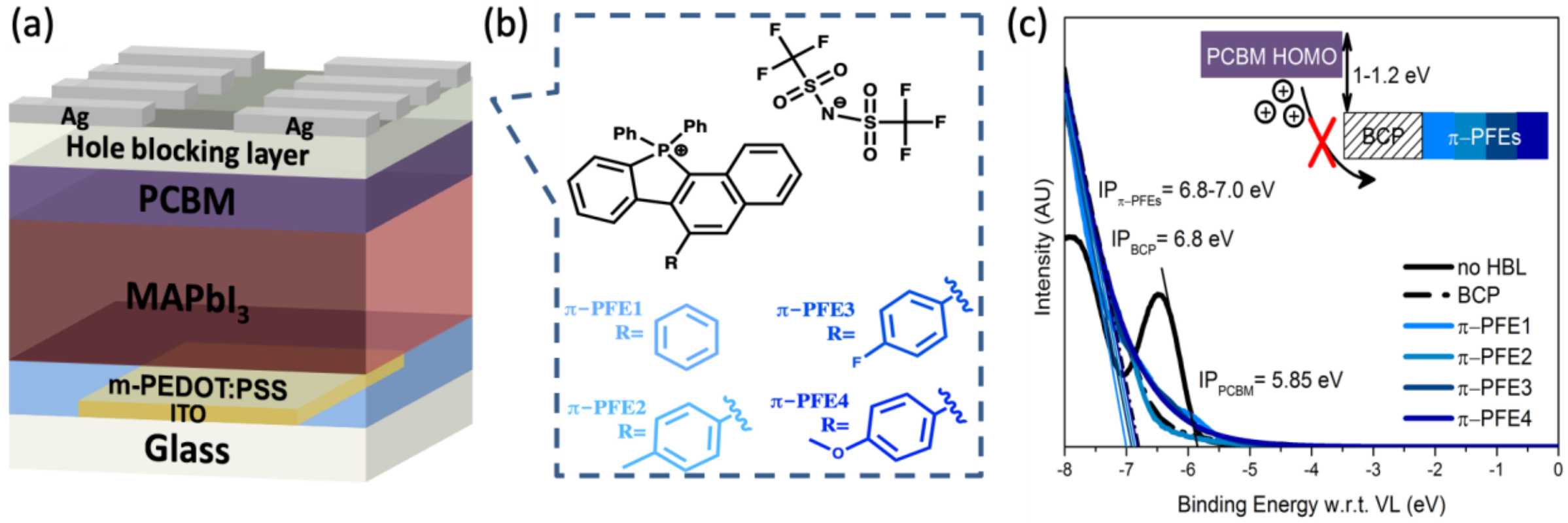

**Figure 1**: (a) Schematic photovoltaic device structure (b) chemical structure of the four different π-extended phosphoniumfluorenes (c) Ultraviolet photoemission spectroscopy measurements of the ionisation potential of BCP and π-PFE1-4 HBL with PCBM spectrum for reference.

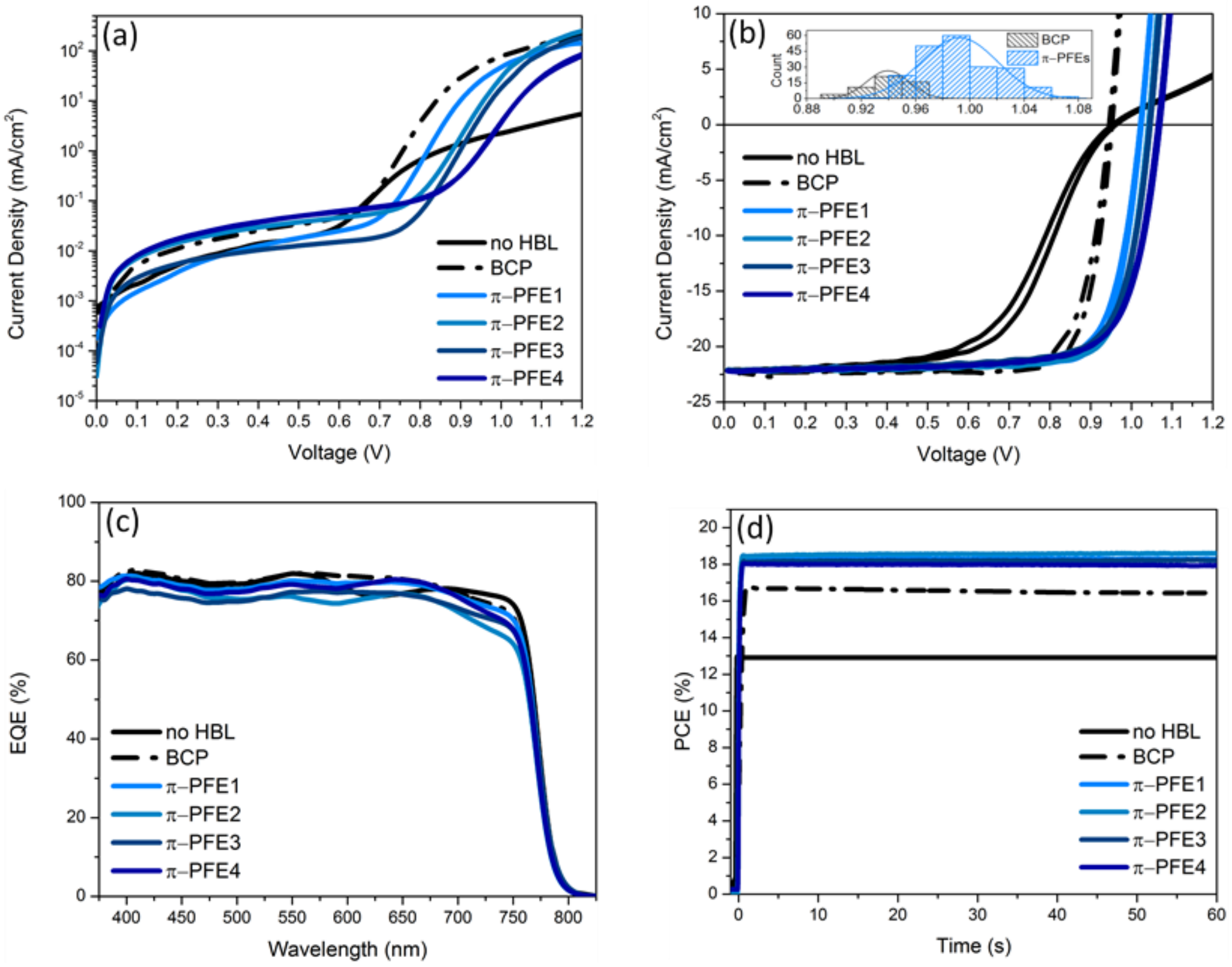

**Figure 2**: Current-voltage measured in (a) dark and (b) under AM 1.5 illumination with corresponding (c) external quantum efficiency (EQE) spectra and (d) time dependence of the maximum power output efficiency of devices with the structure Glass/ITO/PEDOT:PSS/ $MAPbI_3$/PCBM/π-PFE/Ag. Reference devices without a HBL or with a BCP HBL are shown for comparison. The inset in panel (b) shows the histograms of the open-circuit voltages measured on 258 devices with either a BCP or π-PFE HBLs (the complete statistics of the photovoltaic parameters for these devices is shown in **Figure S2**).

**Table 1:** Photovoltaic parameters of optimal $MAPbI_3$ devices with bare PCBM, PCBM/BCP and PCBM/π-PFE1-4. FS and RS represent scanning direction from $J_{SC}$ to $V_{OC}$ and $V_{OC}$ to $J_{SC}$, respectively. $R_S$ is the series resistance of the devices.

| | **$V_{OC}$ FS (V)** | **$J_{SC}$ FS (mA/cm$^2$)** | **FF FS (%)** | **PCE FS (%)** | **$V_{OC}$ RS (V)** | **$J_{SC}$ RS (mA/cm$^2$)** | **FF RS (%)** | **PCE RS (%)** | **$R_S$ (Ω/cm$^2$)** |
|---|---|---|---|---|---|---|---|---|---|
| **PCBM** | 0.95 | -22.22 | 56.84 | 12.03 | 0.96 | -22.22 | 60.1 | 12.85 | 33.26 |
| **BCP** | 0.93 | -22.40 | 78.32 | 16.33 | 0.95 | -22.40 | 81.39 | 17.38 | 2.90 |
| **π-PFE1** | 1.02 | -22.19 | 78.23 | 17.74 | 1.03 | -22.19 | 80.55 | 18.34 | 2.33 |
| **π-PFE2** | 1.04 | -22.11 | 78.25 | 18.03 | 1.04 | -22.11 | 79.90 | 18.46 | 2.22 |
| **π-PFE3** | 1.05 | -22.05 | 76.90 | 17.72 | 1.05 | -22.05 | 78.09 | 18.05 | 2.20 |
| **π-PFE4** | 1.07 | -22.16 | 76.07 | 17.96 | 1.07 | -22.16 | 75.95 | 18.03 | 2.40 |

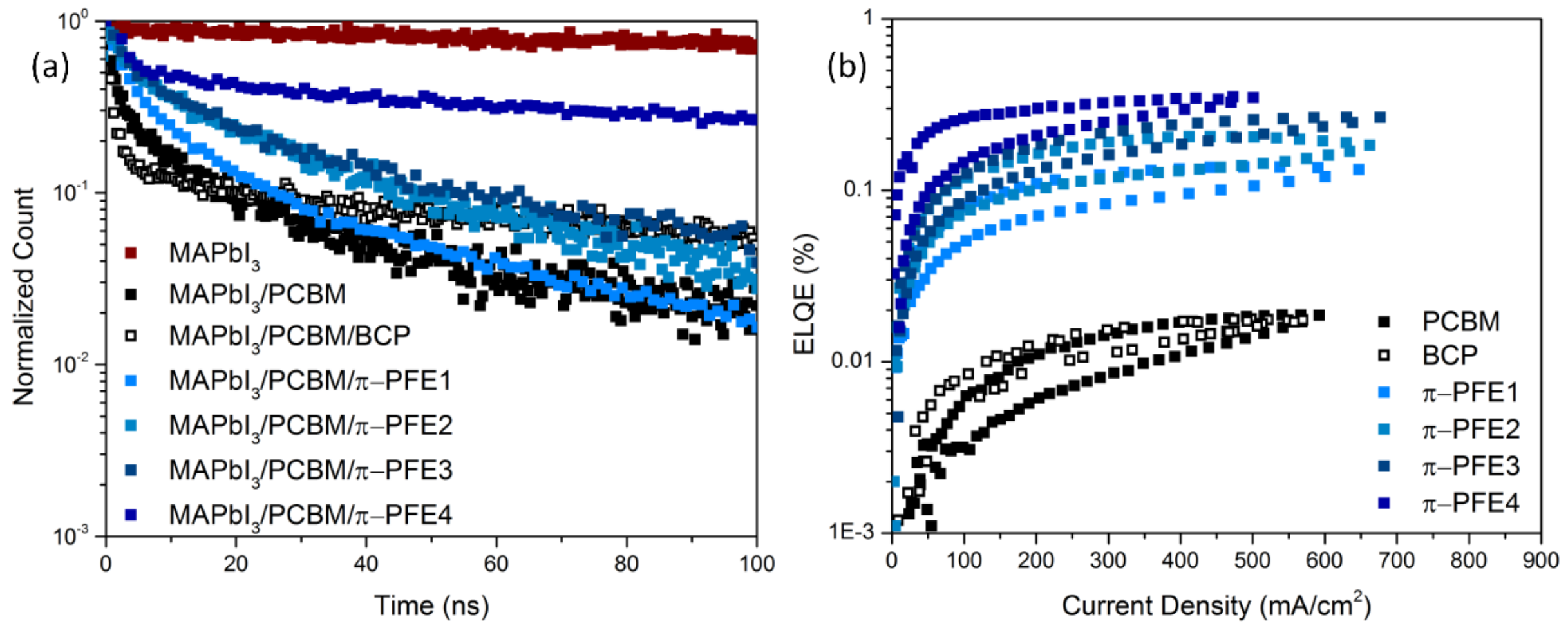

**Figure 3**: (a) Normalized time correlated single photon counting (TCSPC) photon decay for films of perovskite only, perovskite/PCBM and perovskite/PCBM/HBLs on glass substrates. (b) Electroluminescence quantum efficiency of devices with and without HBLs.

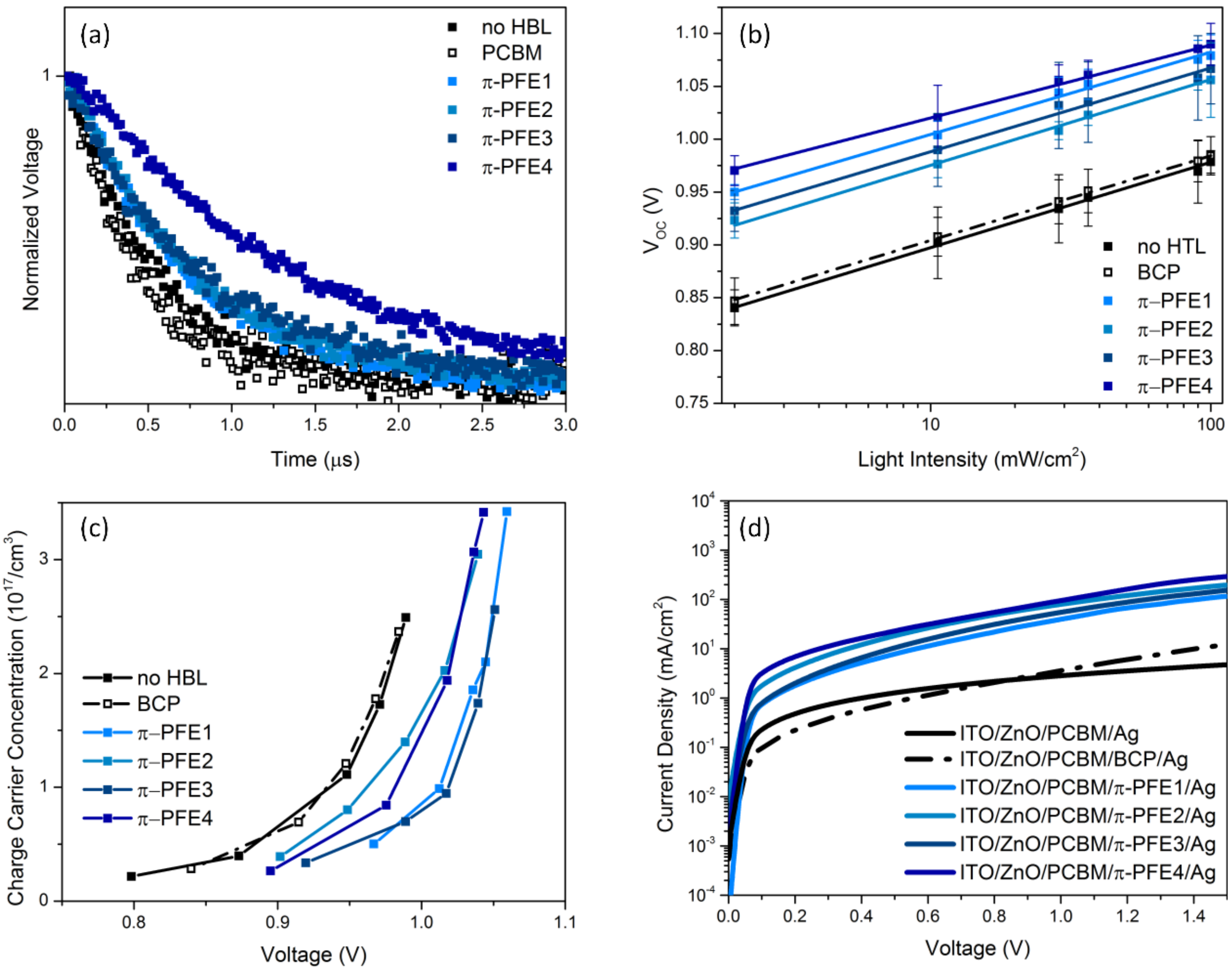

**Figure 4**: (a) Transient photovoltage measurements of devices with and without HBLs (b) $V_{OC}$ dependence upon different light intensity of perovskite solar cell with PCBM only and with PCBM/HBLs. (c) Charge density generated in the device with and without HBLs, measuring by transient photovoltage and transient photocurrent methods. (d) J-V characteristics of unipolar electron only devices.

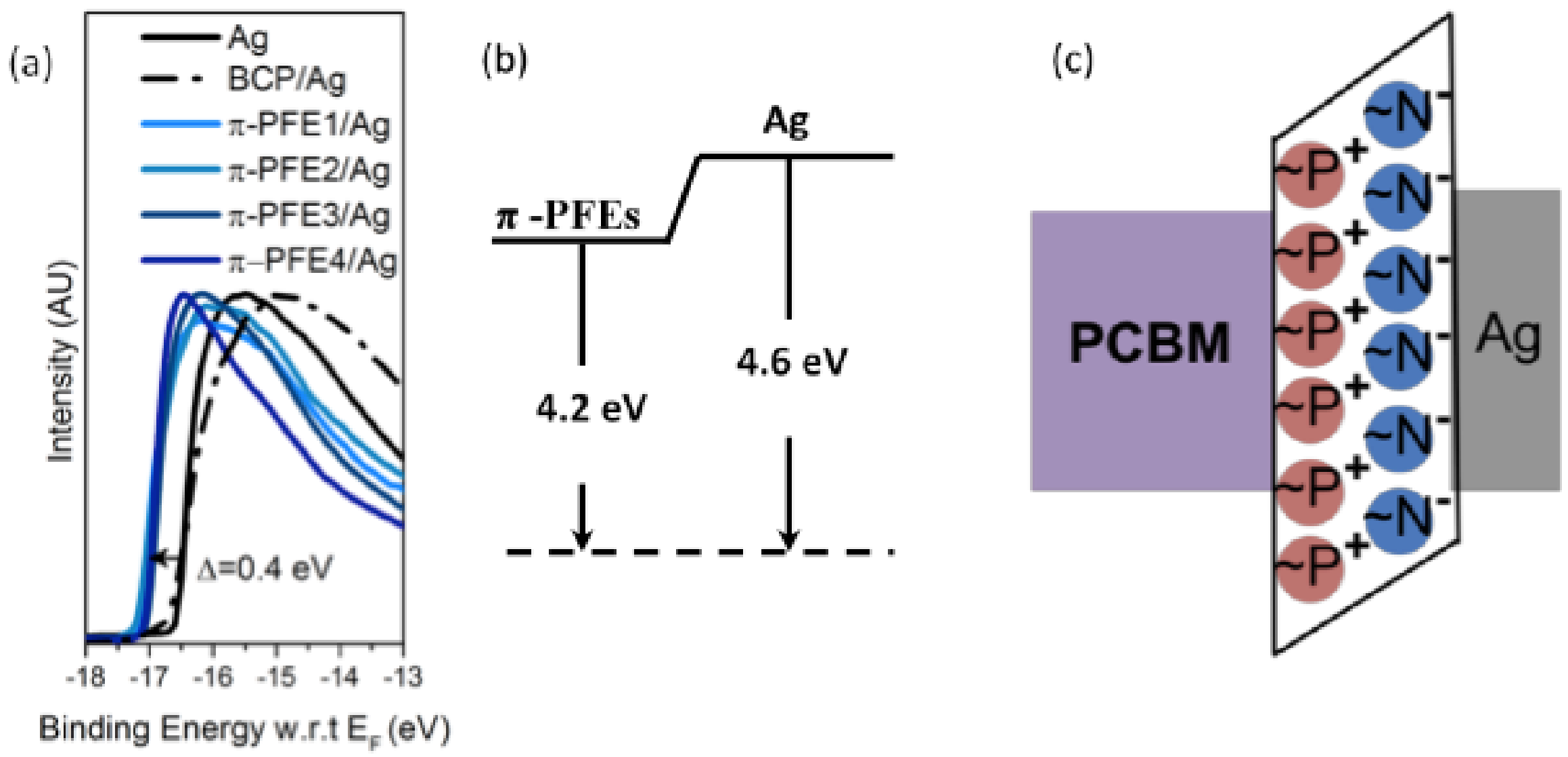

**Figure 5:** Interfacial dipole of π-extended phosphoniumfluorenes (1-4) on Ag. (a) UPS spectrum of Ag (evaporated), BCP/Ag and π-PFE1-4/Ag (detached from ITO/PCBM/HBL/Ag devices) with secondary electron cut-off and HOMO region. (b) Energy level diagram obtained from (a). (c) Schematic illustration of dipole formed by π-PFE between PCBM/Ag.

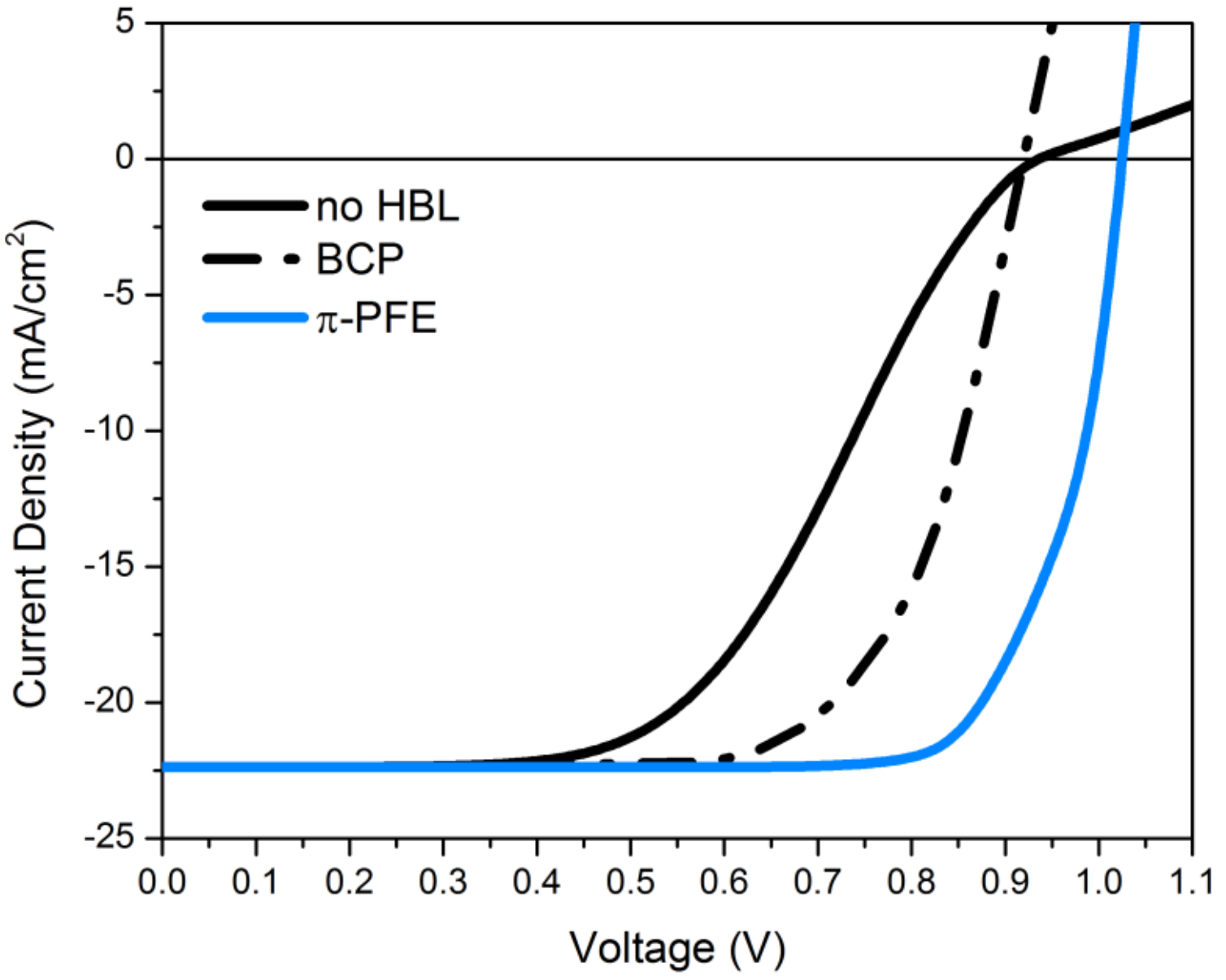

**Figure 6**: Simulated stabilized current−voltage characteristics under one sun for either no HBL, BCP or π-PFE HBL.